\documentclass{revtex4}
\usepackage{amsmath}
\usepackage{amssymb}
\usepackage{graphicx}

 \evensidemargin0in \oddsidemargin0in \topmargin10pt
\begin{document}

\title{Wigner functions of thermo number state, photon subtracted and added
thermo vacuum state at finite temperature{\small \thanks{%
Project supported by the National Natural Science Foundation of China (Grant
Nos 10775097 and 10874174).}}}
\author{Li-yun Hu {\small \thanks{%
Corresponding author. \emph{E-mail address}: hlyun2008@126.com (L-Y Hu).}}
and Hong-yi Fan \\
Department of Physics, Shanghai Jiao Tong University, Shanghai 200030, China}

\begin{abstract}
Based on Takahashi-Umezawa thermo field dynamics and the order-invariance of
Weyl ordered operators under similar transformations, we present a new
approach to deriving the exact Wigner functions of thermo number state,
photon subtracted and added thermo vacuum state. We find that these Wigner
functions are related to the Gaussian-Laguerre type functions of
temperature, whose statistical properties are then analysed.
\end{abstract}
\maketitle
\section{Introduction}

In recent years photon subtracted and added quantum states have been paid
much attention because these fields exhibit an abundant of nonclassical
properties\ and may give access to a complete engineering of quantum states
and to fundamental quantum phenomena [1-8]. However, all these discussions
are restricted to the case at zero point temperature. In fact, most systems
are not isolated, but are immersed in a \textquotedblleft thermal
reservoir", excitation and de-excitation processes of a system are
influenced by its energy exchange with reservoirs.\ In this work we study
field properties by photon subtracting and adding at finite temperature.

The Wigner function (WF) is a powerful tool to investigate the
nonclassicality of optical fields [9,10]. Its partial negativity implies the
highly nonclassical properties of quantum states and is often used to
describe the decoherence of quantum states [7,8,11,12]. In one dimensional
case, the WF of a density matrix $\rho $ is defined as $\mathtt{Tr}\left[
\rho \Delta (\alpha )\right] ,$ where $\Delta (\alpha )$ is the single-mode
Wigner operator, whose normally ordered form and Weyl ordered form are given
as [13-15], respectively,
\begin{equation}
\Delta \left( \alpha \right) =\frac{1}{\pi }\colon e^{-\left( q-Q\right)
^{2}-\left( p-P\right) ^{2}}\colon =\frac{1}{\pi }\colon e^{-2\left( \alpha
-a\right) \left( \alpha ^{\ast }-a^{\dagger }\right) }\colon ,  \label{1}
\end{equation}%
and%
\begin{equation}
\Delta \left( \alpha \right)
=\frac{1}{2}\genfrac{}{}{0pt}{}{:}{:}\delta \left( \alpha -a\right)
\delta \left( \alpha ^{\ast }-a^{\dagger }\right)
\genfrac{}{}{0pt}{}{:}{:}, \label{2}
\end{equation}%
where $\alpha =\left( q+\mathtt{i}p\right) /\sqrt{2},$ $a=\left( Q+\mathtt{i}%
P\right) /\sqrt{2}$, $\left[ Q,P\right] =\mathtt{i},$ $\hbar =1;$ $a$ and $%
a^{\dagger }$ ($\left[ a,a^{\dagger }\right] =1)$ are Bose annihilation and
creation operators, the symbols $\colon \colon $ and $\genfrac{}{}{0pt}{}{:}{:}\genfrac{}{}{0pt}{}{:}{:%
}$ denote the normal ordering and the Weyl ordering, respectively.
Our main aim is to provide a new and direct approach to deriving the
WFs of quantum states at finite temperature by using the
order-invariance of Weyl ordered
operators under similar transformations [13-15], which means%
\begin{equation}
S\genfrac{}{}{0pt}{}{:}{:}\left( \circ \circ \circ \right) \genfrac{}{}{0pt}{}{:}{:}S^{-1}=\genfrac{}{}{0pt}{}{:}{:}%
S\left( \circ \circ \circ \right) S^{-1}\genfrac{}{}{0pt}{}{:}{:},
\label{2a}
\end{equation}%
as if the \textquotedblleft fence"
$\genfrac{}{}{0pt}{}{:}{:}\genfrac{}{}{0pt}{}{:}{:}$did not exist,
so $S$ can pass through it. We also appeal to the Takahashi-Umezawa
thermo field dynamics (TFD) [16-18], we consider it convenient to
obtaining the explicit expressions of WFs.

\section{Brief review of thermo state}

The main point of TFD lies in converting the evaluation of ensemble average
at nonzero temperature into the equivalent expectation value with a pure
state. This worthwhile convenience is at the expense of introducing a
fictitious field (or a so-called tilde-conjugate field, denoted as operator $%
\tilde{a}^{\dagger }$) in the extending Hilbert space $\tilde{H}$, thus the
original optical field state $\left\vert n\right\rangle $\ in the Hilbert
space $\mathcal{H}$\ is accompanied by a tilde state $\left\vert \tilde{n}%
\right\rangle $\ in $\tilde{H}$. \ A similar rule holds for operators: every
annihilation operator $a$\ acting on $\mathcal{H}$\ has an image $\tilde{a}$%
\ acting on $\tilde{H}$. At finite temperature $T$ the thermal vacuum $%
\left\vert 0(\beta )\right\rangle $ is defined by the requirement that the
vacuum expectation value agrees with the statistical average [16-18], i.e.%
\begin{equation}
\left\langle A\right\rangle =\mathtt{Tr}\left( \rho _{c}A\right)
=\left\langle 0(\beta )\right\vert A\left\vert 0(\beta )\right\rangle =%
\mathtt{Tr}\left( Ae^{-\beta H}\right) /\mathtt{Tr}\left( e^{-\beta
H}\right) ,  \label{3}
\end{equation}%
where $\beta =\frac{1}{kT},$ $k$ is the Boltzmann constant and $H$ is the
system's Hamiltonian. For the ensemble of free bosons with Hamiltonian $%
H_{0}=\omega a^{\dag }a$, the thermal vacuum state $\left\vert 0(\beta
)\right\rangle $\ is%
\begin{equation}
\left\vert 0(\beta )\right\rangle =\text{sech}\theta \exp \left[ a^{\dagger }%
\tilde{a}^{\dagger }\tanh \theta \right] \left\vert 0,\tilde{0}\right\rangle
=S\left( \theta \right) \left\vert 0,\tilde{0}\right\rangle ,  \label{4}
\end{equation}%
where $\left\vert 0,\tilde{0}\right\rangle $ is annihilated by $a$ and $%
\tilde{a},$ $\left[ \tilde{a},\tilde{a}^{\dagger }\right] =1,$ and
\begin{equation}
S\left( \theta \right) \equiv \exp \left[ \theta \left( a^{\dagger }\tilde{a}%
^{\dagger }-a\tilde{a}\right) \right] ,  \label{5}
\end{equation}%
is the thermo squeezing operator which transforms the zero-temperature
vacuum $\left\vert 0,\tilde{0}\right\rangle $ into the thermo vacuum state $%
\left\vert 0(\beta )\right\rangle ,$ and $\theta $ is related to the Bose
distribution by
\begin{equation}
\tanh \theta =\exp \left( -\frac{\omega }{2kT}\right) ,  \label{6}
\end{equation}%
which is determined by comparing the Bose--Einstein distribution%
\begin{equation}
n_{c}=\left[ \exp \left( \frac{\omega }{kT}\right) -1\right] ^{-1}  \label{7}
\end{equation}%
and%
\begin{equation}
\left\langle 0(\beta )\right\vert a^{\dagger }a\left\vert 0(\beta
)\right\rangle =\sinh ^{2}\theta .  \label{8}
\end{equation}%
In particular, when operator $A$ is the Wigner operator $\Delta \left(
\alpha \right) $ itself, it is easy to see that
\begin{eqnarray}
\mathtt{Tr}_{a}\left( \Delta \left( \alpha \right) e^{-\beta H}\right) /%
\mathtt{Tr}_{a}\left( e^{-\beta H}\right)  &=&\left\langle 0(\beta
)\right\vert \Delta \left( \alpha \right) \left\vert 0(\beta )\right\rangle
\notag \\
&=&\mathtt{Tr}_{a,\tilde{a}}\left[ \Delta \left( \alpha \right) \left\vert
0(\beta )\right\rangle \left\langle 0(\beta )\right\vert \right] ,  \label{9}
\end{eqnarray}%
which is just the WF of thermo vacuum state. From Eq.(\ref{9}) one can see
that the calculation of WF for thermo states is converted into the
expectation value of Wigner operator in themo vacuum state $\left\vert
0(\beta )\right\rangle $ ($\rho _{c}\rightarrow \left\vert 0(\beta
)\right\rangle \left\langle 0(\beta )\right\vert $), which is defined in the
enlarged\ Fock space. This implies that it is convenient to deriving some
WFs of density operators at finite temperature by doubly enlarging the
original space.

\section{Normally ordered form of $S^{\dagger }\left( \protect\theta \right)
\Delta \left( \protect\alpha \right) S\left( \protect\theta \right) $}

In order to deriving conveniently the WFs of density operators at finite
temperature, let's first calculate the normally ordered form of $S^{\dagger
}\left( \theta \right) \Delta \left( \alpha \right) S\left( \theta \right) .$
Recalling that for single-mode case the Weyl rule [13-15] is defined as%
\begin{equation}
\hat{H}\left( a,a^{\dag }\right) =2\int \mathtt{d}^{2}\alpha h\left( \alpha
,\alpha ^{\ast }\right) \Delta \left( \alpha \right) ,  \label{10}
\end{equation}%
where $h\left( \alpha ,\alpha ^{\ast }\right) $ is the classical function
corresponding to operator $\hat{H}\left( a,a^{\dag }\right) .$ Eq.(\ref{10})
expresses the Weyl correspondence rule, using (\ref{2}) it can be expressed
as%
\begin{eqnarray}
\hat{H}\left( a,a^{\dag }\right)  &=&\int \mathtt{d}^{2}\alpha
h\left( \alpha ,\alpha ^{\ast }\right)
\genfrac{}{}{0pt}{}{:}{:}\delta \left( \alpha -a\right)
\delta \left( \alpha ^{\ast }-a^{\dagger }\right) \genfrac{}{}{0pt}{}{:}{:}  \notag \\
&=&\genfrac{}{}{0pt}{}{:}{:}h\left( a,a^{\dagger }\right)
\genfrac{}{}{0pt}{}{:}{:}, \label{11}
\end{eqnarray}%
which means that Weyl ordered of operator
$\genfrac{}{}{0pt}{}{:}{:}h\left( a,a^{\dagger }\right)
\genfrac{}{}{0pt}{}{:}{:}$, whose Weyl correspondence is\ $h\left(
\alpha
,\alpha ^{\ast }\right) $, can be obtained by just respectively replacing $%
\alpha ,\alpha ^{\ast }$ in $h\left( \alpha ,\alpha ^{\ast }\right) $ by $a$
and $a^{\dagger }$ without disturbing the form of function $h$.

According to the Weyl ordering invariance under similar transformations [13]
and the following transform relation
\begin{eqnarray}
S^{\dagger }\left( \theta \right) aS\left( \theta \right)  &=&a\cosh \theta +%
\tilde{a}^{\dagger }\sinh \theta ,  \notag \\
S^{\dagger }\left( \theta \right) \tilde{a}S\left( \theta \right)  &=&\tilde{%
a}\cosh \theta +a^{\dagger }\sinh \theta ,  \label{12}
\end{eqnarray}%
it is easily seen%
\begin{eqnarray}
S^{\dagger }\left( \theta \right) \Delta \left( \alpha \right)
S\left( \theta \right)
&=&\frac{1}{2}\genfrac{}{}{0pt}{}{:}{:}\delta \left( \alpha -a\cosh
\theta -\tilde{a}^{\dagger }\sinh \theta \right)   \notag \\
&&\times \delta \left( \alpha ^{\ast }-a^{\dag }\cosh \theta
-\tilde{a}\sinh \theta \right) \genfrac{}{}{0pt}{}{:}{:},
\label{13}
\end{eqnarray}%
which is just the Weyl ordering of $S^{\dagger }\left( \theta \right) \Delta
\left( \alpha \right) S\left( \theta \right) $ in the enlarged Fock space.
Based on the Weyl rule, the classical correspondence $h\left( \beta ,\beta
^{\ast };\tilde{\beta},\tilde{\beta}^{\ast }\right) $ of the operator $%
S^{\dagger }\left( \theta \right) \Delta \left( \alpha \right) S\left(
\theta \right) $ can be obtained by replacing ($a,a^{\dag })$ and ($\tilde{a}%
,\tilde{a}^{\dag })$ with ($\beta ,\beta ^{\ast }$)$\ $and ($\tilde{\beta},%
\tilde{\beta}^{\ast })$, respectively, i.e.,
\begin{eqnarray}
h\left( \beta ,\beta ^{\ast };\tilde{\beta},\tilde{\beta}^{\ast }\right)  &=&%
\frac{1}{2}\delta \left( \alpha -\beta \cosh \theta -\tilde{\beta}^{\ast
}\sinh \theta \right)   \notag \\
&&\times \delta \left( \alpha ^{\ast }-\beta ^{\ast }\cosh \theta -\tilde{%
\beta}\sinh \theta \right) .  \label{14}
\end{eqnarray}%
It then follows from Eqs.(\ref{10}) and (\ref{14}) that%
\begin{equation}
S^{\dagger }\left( \theta \right) \Delta \left( \alpha \right) S\left(
\theta \right) =4\int \mathtt{d}^{2}\beta \mathtt{d}^{2}\tilde{\beta}\Delta
\left( \beta ,\beta ^{\ast };\tilde{\beta},\tilde{\beta}^{\ast }\right)
h\left( \beta ,\beta ^{\ast };\tilde{\beta},\tilde{\beta}^{\ast }\right) ,
\label{15}
\end{equation}%
where $\Delta \left( \beta ,\beta ^{\ast };\tilde{\beta},\tilde{\beta}^{\ast
}\right) $ is the two-mode Wigner operator, whose normally ordering form is%
\begin{equation}
\Delta \left( \beta ,\beta ^{\ast };\tilde{\beta},\tilde{\beta}^{\ast
}\right) =\frac{1}{\pi ^{2}}\colon \exp \left[ -2\left( a^{\dag }-\beta
^{\ast }\right) \left( a-\beta \right) -2\left( \tilde{a}^{\dag }-\tilde{%
\beta}^{\ast }\right) \left( \tilde{a}-\tilde{\beta}\right) \right] \colon .
\label{16}
\end{equation}%
On substituting Eq.(\ref{16}) into Eq.(\ref{15}) and using the integral
formula [19]%
\begin{equation}
\int \frac{\mathtt{d}^{2}z}{\pi }e^{\zeta \left\vert z\right\vert ^{2}+\xi
z+\eta z^{\ast }}=-\frac{1}{\zeta }e^{-\frac{\xi \eta }{\zeta }},\text{ Re}%
\left( \zeta \right) <0,  \label{17}
\end{equation}%
we can derive the normally ordered form of (\ref{15}) as follows%
\begin{eqnarray}
S^{\dagger }\left( \theta \right) \Delta \left( \alpha \right) S\left(
\theta \right)  &=&2\int \frac{\mathtt{d}^{2}\beta \mathtt{d}^{2}\tilde{\beta%
}}{\pi ^{2}}\delta \left( \alpha -\beta \cosh \theta -\tilde{\beta}^{\ast
}\sinh \theta \right)   \notag \\
&&\times \delta \left( \alpha ^{\ast }-\beta ^{\ast }\cosh \theta -\tilde{%
\beta}\sinh \theta \right)   \notag \\
&&\times \colon \exp \left[ -2\left( a^{\dag }-\beta ^{\ast }\right) \left(
a-\beta \right) -2\left( \tilde{a}^{\dag }-\tilde{\beta}^{\ast }\right)
\left( \tilde{a}-\tilde{\beta}\right) \right] \colon   \notag \\
&=&\frac{\text{sech}2\theta }{\pi }e^{-2\left\vert \alpha \right\vert ^{2}%
\text{sech}2\theta }\colon \exp \left\{ -\left( a\tilde{a}+a^{\dag }\tilde{a}%
^{\dag }\right) \tanh 2\theta \right.   \notag \\
&&+2\text{sech}2\theta \left[ \sinh \theta \left( \allowbreak \alpha ^{\ast }%
\tilde{a}^{\dag }+\allowbreak \alpha \tilde{a}\right) +\cosh \theta \left(
\alpha ^{\ast }a+\alpha a^{\dagger }\right) \right.   \notag \\
&&-\left( \tilde{a}^{\dag }\tilde{a}\sinh ^{2}\theta +a^{\dag }a\cosh
^{2}\theta \right) ]\}\colon ,  \label{18}
\end{eqnarray}%
which is just the normally ordered form of (\ref{15}). Eq.(\ref{18})
directly leads to the WF of the thermo vacuum state $\left\vert 0(\beta
)\right\rangle $,
\begin{eqnarray}
\left\langle 0(\beta )\right\vert \Delta \left( \alpha \right) \left\vert
0(\beta )\right\rangle  &=&\left\langle 0,\tilde{0}\right\vert S^{\dagger
}\left( \theta \right) \Delta \left( \alpha \right) S\left( \theta \right)
\left\vert 0,\tilde{0}\right\rangle =\frac{\text{sech}2\theta }{\pi }%
e^{-2\left\vert \alpha \right\vert ^{2}\text{sech}2\theta }  \notag \\
&=&\frac{1-e^{-\beta \omega }}{\pi (1+e^{-\beta \omega })}e^{-2\left\vert
\alpha \right\vert ^{2}\frac{1-e^{-\beta \omega }}{1+e^{-\beta \omega }}}.
\label{19}
\end{eqnarray}

\section{Wigner function of photon-subtracted thermo vacuum state}

At finite temperature, the photon-subtracted thermo vacuum state can be
expressed as [20]

\begin{equation}
\rho _{1}=C_{1}a^{n}\left\vert 0(\beta )\right\rangle \left\langle 0(\beta
)\right\vert a^{\dag n},  \label{20}
\end{equation}%
where $C_{1}$ is the normalized factor, defined by%
\begin{equation}
C_{1}^{-1}=\mathtt{Tr}\left[ a^{n}S\left( \theta \right) \left\vert 0,\tilde{%
0}\right\rangle \left\langle 0,\tilde{0}\right\vert S^{\dag }\left( \theta
\right) a^{\dag n}\right] ,  \label{21}
\end{equation}%
which can be calculated as follows. Using Eq.(\ref{4}) and the binomial
formula%
\begin{equation}
\sum_{l=0}^{\infty }\frac{\left( n+l\right) !}{n!l!}x^{l}=\left( 1-x\right)
^{-n-1},  \label{22}
\end{equation}%
we have%
\begin{eqnarray}
C_{1}^{-1} &=&\left\langle 0,\tilde{0}\right\vert S^{\dag }\left( \theta
\right) a^{\dag n}a^{n}S\left( \theta \right) \left\vert 0,\tilde{0}%
\right\rangle   \notag \\
&=&\text{sech}^{2}\theta \left\langle 0,\tilde{0}\right\vert e^{a\tilde{a}%
\tanh \theta }a^{\dag n}a^{n}e^{a^{\dagger }\tilde{a}^{\dagger }\tanh \theta
}\left\vert 0,\tilde{0}\right\rangle   \notag \\
&=&\text{sech}^{2}\theta \sum_{k,l=0}^{\infty }\tanh ^{l+k}\theta
\left\langle k,\tilde{k}\right\vert a^{\dag n}a^{n}\left\vert l,\tilde{l}%
\right\rangle   \notag \\
&=&\text{sech}^{2}\theta \sum_{l=n}^{\infty }\frac{l!}{\left( l-n\right) !}%
\tanh ^{2l}\theta =n!\sinh ^{2n}\theta .  \label{23}
\end{eqnarray}%
By using Eqs. (\ref{20}) and (\ref{18}), we calculate the WF of
photon-subtracted thermal state $\rho _{1}$
\begin{eqnarray}
W_{1}\left( \alpha \right)  &=&C_{1}\left\langle 0,\tilde{0}\right\vert
S^{\dag }\left( \theta \right) a^{\dag n}\Delta \left( \alpha \right)
a^{n}S\left( \theta \right) \left\vert 0,\tilde{0}\right\rangle   \notag \\
&=&\left\langle 0,\tilde{0}\right\vert \left[ S^{\dag }\left( \theta \right)
a^{\dag n}S\left( \theta \right) \right] S^{\dag }\left( \theta \right)
\Delta \left( \alpha \right) S\left( \theta \right) \left[ S^{\dag }\left(
\theta \right) a^{n}S\left( \theta \right) \right] \left\vert 0,\tilde{0}%
\right\rangle .  \label{24}
\end{eqnarray}%
Noticing Eq.(\ref{12}) we see
\begin{eqnarray}
\left[ S^{\dag }\left( \theta \right) a^{n}S\left( \theta \right) \right]
\left\vert 0,\tilde{0}\right\rangle  &=&\left( a\cosh \theta +\tilde{a}%
^{\dagger }\sinh \theta \right) ^{n}\left\vert 0,\tilde{0}\right\rangle
\notag \\
&=&\sqrt{n!}\sinh ^{n}\theta \left\vert 0,\tilde{n}\right\rangle ,
\label{25}
\end{eqnarray}%
then substituting (\ref{25}) into Eq.(\ref{24}) and using Eq.(\ref{18})
yields%
\begin{eqnarray}
W_{1}\left( \alpha \right)  &=&\frac{e^{-2\left\vert \alpha \right\vert ^{2}%
\text{sech}2\theta }}{\pi \cosh 2\theta }\left\langle \tilde{n}\right\vert
e^{\frac{2\sinh \theta }{\cosh 2\theta }\alpha ^{\ast }\tilde{a}^{\dag
}}\left( \text{sech}2\theta \right) ^{\tilde{a}^{\dag }\tilde{a}}e^{\frac{%
2\sinh \theta }{\cosh 2\theta }\tilde{a}\alpha }\left\vert \tilde{n}%
\right\rangle   \notag \\
&=&\frac{e^{-2\left\vert \alpha \right\vert ^{2}\text{sech}2\theta }}{\pi
\cosh 2\theta }\sum_{k,l=0}^{n}\frac{\alpha ^{\ast k}\alpha ^{l}}{k!l!}%
\left( \frac{2\sinh \theta }{\cosh 2\theta }\right) ^{k+l}\left\langle
\tilde{n}\right\vert \tilde{a}^{\dag k}\left( \text{sech}2\theta \right) ^{%
\tilde{a}^{\dag }\tilde{a}}\tilde{a}^{l}\left\vert \tilde{n}\right\rangle
\notag \\
&=&\frac{e^{-2\left\vert \alpha \right\vert ^{2}\text{sech}2\theta }}{\pi
\cosh ^{n+1}2\theta }\sum_{l=0}^{n}\frac{n!}{l!l!\left( n-l\right) !}\left(
\frac{4\sinh ^{2}\theta }{\cosh 2\theta }\left\vert \alpha \right\vert
^{2}\right) ^{l},  \label{26}
\end{eqnarray}%
where we have used the identity operator [21]%
\begin{equation}
\exp \left[ \lambda \tilde{a}^{\dag }\tilde{a}\right] =\colon \exp \left[
\left( e^{\lambda }-1\right) \tilde{a}^{\dag }\tilde{a}\right] \colon .
\label{26a}
\end{equation}%
Recalling that the definition of Laguerre polynomials [22],
\begin{equation}
L_{n}(x)=\sum_{l=0}^{n}\frac{n!}{\left( l!\right) ^{2}\left( n-l\right) !}%
(-x)^{l},  \label{27}
\end{equation}%
Eq. (\ref{26}) can be further put into the following neat form,%
\begin{equation}
W_{1}\left( \alpha \right) =\frac{e^{-2\left\vert \alpha \right\vert ^{2}%
\text{sech}2\theta }}{\pi \cosh ^{n+1}2\theta }L_{n}\left( -\frac{4\sinh
^{2}\theta }{\cosh 2\theta }\left\vert \alpha \right\vert ^{2}\right) ,
\label{28}
\end{equation}%
which is just the WF of photon-subtracted thermo vacuum state, a
Gaussian-Laguerre type function of temperature, since $\tanh \theta =\exp
\left( -\frac{\omega }{2kT}\right) $. Due to $\cosh 2\theta >0$ and $L_{n}(-%
\frac{4\sinh ^{2}\theta }{\cosh 2\theta }\left\vert \alpha \right\vert
^{2})\geqslant 0$, for the photon-subtracted case, $W_{1}\left( \alpha
\right) $ has no chance to present the negative value in phase space, which
can be seen from Fig.1. On the other hand, the amplitude value of WF in $%
\left( \left\vert \alpha \right\vert ,\theta \right) $ space decreases with
the increasing temperature (corresponding to $\theta $). In appendix A, in
order to check the result in Eq. (\ref{28}), we have derived the WF of
photon-subtracted thermo vacuum state by using the coherent state
representation of Wigner operator. Comparing with the result in Ref.[20],
Eq.(\ref{28}) seems \emph{more concise and convenient} for further
discussion.

\begin{figure}[tbp]
\label{Fig1} \centering\includegraphics[width=14cm]{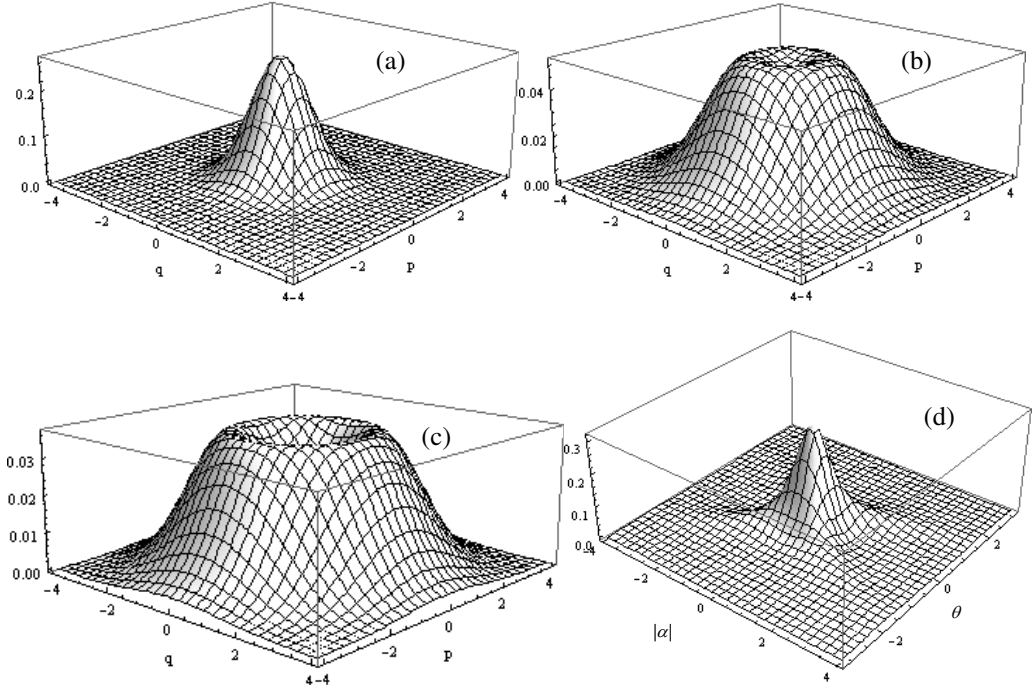}
\caption{Wigner function distributions of photon-subtracted thermo state in (%
$q,p$) phase space with (a) $n=1,\protect\theta =0.2$, (b) $n=1,\protect%
\theta =0.8$, (c) $n=2,\protect\theta =0.8$, and in $\left( \left\vert
\protect\alpha \right\vert ,\protect\theta \right) $ space with (d) $n=1$. }
\end{figure}

\section{Wigner function of photon-added thermo vacuum state}

At finite temperature, the photon-added thermo vacuum state is expressed as
[23]

\begin{equation}
\rho _{2}=C_{2}a^{\dag n}\left\vert 0(\beta )\right\rangle \left\langle
0(\beta )\right\vert a^{n}.  \label{29}
\end{equation}%
By the same procedures as deriving Eqs. (\ref{21}) and (\ref{25}), we have%
\begin{equation}
C_{2}^{-1}=n!\cosh ^{2n}\theta ,  \label{30}
\end{equation}%
and
\begin{equation}
S^{\dag }\left( \theta \right) a^{\dag n}S\left( \theta \right) \left\vert 0,%
\tilde{0}\right\rangle =\sqrt{n!}\cosh ^{n}\theta \left\vert n,\tilde{0}%
\right\rangle .  \label{31}
\end{equation}%
Uisng Eq.(\ref{30}) and (\ref{31}), the WF $W_{2}\left( \alpha \right) $ of $%
\rho _{2}\ $is given by%
\begin{eqnarray}
W_{2}\left( \alpha \right) &=&C_{2}\left\langle 0,\tilde{0}\right\vert
S^{\dag }\left( \theta \right) a^{n}S\left( \theta \right) \left[ S^{\dag
}\left( \theta \right) \Delta \left( \alpha \right) S\left( \theta \right) %
\right] S^{\dag }\left( \theta \right) a^{\dag n}S\left( \theta \right)
\left\vert 0,\tilde{0}\right\rangle  \notag \\
&=&\left\langle n,\tilde{0}\right\vert S^{\dag }\left( \theta \right) \Delta
\left( \alpha \right) S\left( \theta \right) \left\vert n,\tilde{0}%
\right\rangle  \notag \\
&=&\frac{\left( -1\right) ^{n}e^{-2\left\vert \alpha \right\vert ^{2}\text{%
sech}2\theta }}{\pi \cosh ^{n+1}2\theta }\sum_{l=0}^{n}\frac{n!}{l!l!\left(
n-l\right) !}\left( -\frac{4\cosh ^{2}\theta }{\cosh 2\theta }\left\vert
\alpha \right\vert ^{2}\right) ^{l}  \notag \\
&=&\frac{\left( -1\right) ^{n}e^{-2\left\vert \alpha \right\vert ^{2}\text{%
sech}2\theta }}{\pi \cosh ^{n+1}2\theta }L_{n}\left( \frac{4\cosh ^{2}\theta
}{\cosh 2\theta }\left\vert \alpha \right\vert ^{2}\right) ,  \label{32}
\end{eqnarray}%
a Gaussian-Laguerre type function which may present negative region in phase
space (see Fig.2). In particular, when $n=1,$ Eq.(\ref{32}) reduces to%
\begin{equation}
W_{2}\left( \alpha \right) =-\frac{e^{-2\left\vert \alpha \right\vert ^{2}%
\text{sech}2\theta }}{\pi \cosh ^{2}2\theta }\left( 1-\frac{4\cosh
^{2}\theta }{\cosh 2\theta }\left\vert \alpha \right\vert ^{2}\right) .
\label{33}
\end{equation}%
In Fig. 2, the behaviour of WF distributions of photon-added thermo state
are plotted in ($q,p$) phase space and $\left( \left\vert \alpha \right\vert
,\theta \right) $ space. From Fig.2, one can see clearly the modulation
action of photon-added number and temperature. The \textquotedblleft
oscillating frequency" of WF increases with the increasing photon-added
number; while the amplitude value of WF in $\left( \left\vert \alpha
\right\vert ,\theta \right) $ space decreases with the increasing
temperature (corresponding to $\theta $), which indicates that the
nonclassicality is weakened at finite temperature.

\begin{figure}[tbp]
\label{Fig2}\centering\includegraphics[width=14cm]{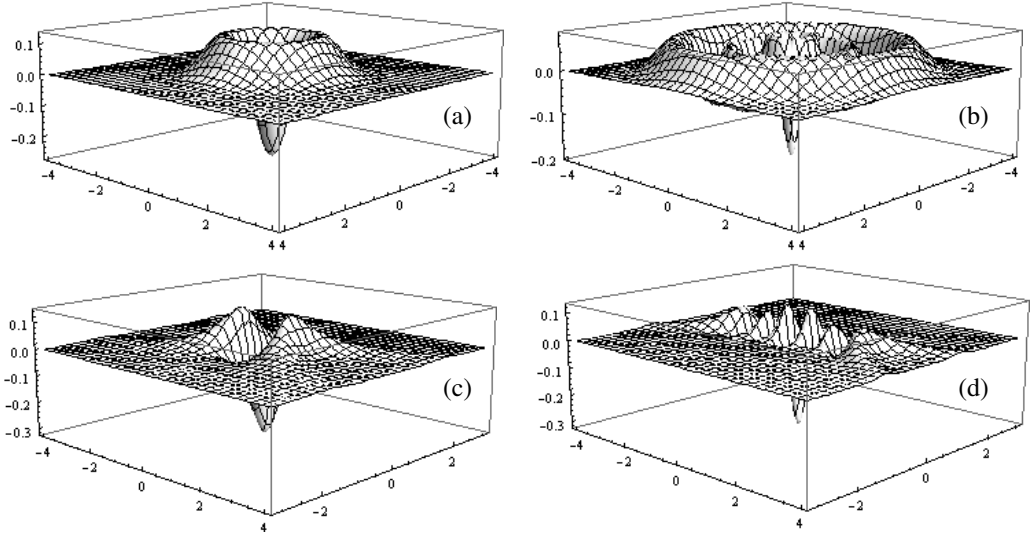}
\caption{Wigner function distributions of photon-added thermo state in ($q,p$%
) phase space with $\protect\theta =0.2$ for (a) $n=1$, (b) $n=2,$ and in $%
\left( \left\vert \protect\alpha \right\vert ,\protect\theta \right) $ space
with (c) $n=1$ and (d) $n=5$.}
\end{figure}

\section{Wigner function of thermo number state}

At finite temperature, according to TFD, the number state $\left\vert
n\right\rangle $ is replaced by $\left\vert n,\tilde{n}\right\rangle ,$ thus
the thermo number state (i.e., number states at finite temperature) is $%
S\left( \theta \right) \left\vert n,\tilde{n}\right\rangle $ in the enlarged
Fock space. Using the un-normalized coherent state representation of number
state,%
\begin{equation}
\left\vert n,\tilde{n}\right\rangle =\frac{1}{n!}\frac{d^{2n}}{dz^{n}d\tilde{%
z}^{n}}\left. \left\vert z,\tilde{z}\right\rangle \right\vert _{z=\tilde{z}%
=0},\text{\ }\left\langle z^{\prime }\right. \left\vert z\right\rangle
=e^{z^{\prime \ast }z},  \label{34}
\end{equation}%
where $\left\vert z,\tilde{z}\right\rangle =\exp [za^{\dag }+\tilde{z}\tilde{%
a}^{\dag }]\left\vert 0,\tilde{0}\right\rangle $ is the non-normalized
two-mode coherent state, and employing Eq.(\ref{18}), we calculate the WF $%
W_{3}\left( \alpha \right) $ of thermo number state as
\begin{eqnarray}
W_{3}\left( \alpha \right)  &=&\left\langle n,\tilde{n}\right\vert S^{\dag
}\Delta \left( \alpha \right) S\left\vert n,\tilde{n}\right\rangle   \notag
\\
&=&\frac{1}{n!^{2}}\frac{d^{2n}}{df^{n}dr^{n}}\frac{d^{2n}}{dz^{n}dt^{n}}%
\left\langle f^{\ast },r^{\ast }\right\vert S^{\dag }\Delta \left( \alpha
\right) S\left. \left\vert z,t\right\rangle \right\vert _{f=r=z=t=0}  \notag
\\
&=&\mathcal{A}\frac{d^{2n}}{df^{n}dr^{n}}\frac{d^{2n}}{dz^{n}dt^{n}}\exp
\left\{ -\left( tz+fr\right) \tanh 2\theta \right.   \notag \\
&&+\left. \left( \allowbreak rt-fz\right) \text{sech}2\theta +zE^{\ast
}+fE+r\allowbreak F^{\ast }+\allowbreak tF\right\} _{f=r=z=t=0},  \label{35}
\end{eqnarray}%
where we have set
\begin{equation}
\mathcal{A=}\frac{e^{-2\left\vert \alpha \right\vert ^{2}\text{sech}2\theta }%
}{\pi n!^{2}\cosh 2\theta },\text{ }E=2\alpha \text{sech}2\theta \cosh
\theta ,\text{ \ }F=2\alpha \text{sech}2\theta \allowbreak \sinh \theta .
\label{36}
\end{equation}%
Expanding the exponential term $\exp \left[ \left( rt-\allowbreak fz\right)
\text{sech}2\theta \right] $ as series, we have%
\begin{eqnarray}
W_{3}\left( \alpha \right)  &=&\mathcal{A}\frac{d^{2n}}{df^{n}dz^{n}}\frac{%
d^{2n}}{dr^{n}dt^{n}}\exp \left[ -\left( fr+tz\right) \tanh 2\theta \right]
\notag \\
&&\times \sum_{l,k=0}^{\infty }\frac{\left( -\allowbreak 1\right) ^{k}\text{%
sech}^{l+k}2\theta }{l!k!}\left( rt\right) ^{l}\left( \allowbreak fz\right)
^{k}\exp \left[ zE^{\ast }+fE+tF\allowbreak +rF^{\ast }\right] _{z=t=f=r=0}
\notag \\
&=&\mathcal{A}\sum_{l,k=0}^{\infty }\frac{\left( -\allowbreak 1\right) ^{k}%
\text{sech}^{l+k}2\theta }{l!k!}\frac{\partial ^{2l}}{\partial
F^{l}\allowbreak \partial F\allowbreak ^{\ast l}}\frac{\partial ^{2k}}{%
\partial E^{k}\allowbreak \partial E\allowbreak ^{\ast k}}  \notag \\
&&\times \frac{d^{2n}}{df^{n}dz^{n}}\frac{d^{2n}}{dr^{n}dt^{n}}\exp \left[
-\left( fr+tz\right) \tanh 2\theta +fE\allowbreak +rF^{\ast }+zE^{\ast }+tF%
\right] _{z=t=f=r=0}.  \label{37}
\end{eqnarray}%
Then making the variable replacement for $f,r,t,z$ we can rewrite Eq.(\ref%
{37}) as

\begin{eqnarray}
W_{3}\left( \alpha \right) &=&\mathcal{A}\tanh ^{2n}2\theta
\sum_{l,k=0}^{\infty }\frac{\left( -\allowbreak 1\right) ^{k}\text{sech}%
^{l+k}2\theta }{l!k!}\frac{\partial ^{2l}}{\partial F^{l}\allowbreak
\partial F\allowbreak ^{\ast l}}\frac{\partial ^{2k}}{\partial
E^{k}\allowbreak \partial E\allowbreak ^{\ast k}}  \notag \\
&&\times \frac{d^{2n}}{df^{n}dr^{n}}\frac{d^{2n}}{dz^{n}dt^{n}}\exp \left[
-fr+fE\allowbreak +\frac{rF^{\ast }}{\tanh 2\theta }-tz+zE^{\ast }+\frac{tF}{%
\tanh 2\theta }\right] _{z=t=f=r=0}  \notag \\
&=&\mathcal{A}\tanh ^{2n}2\theta \sum_{l,k=0}^{\infty }\frac{\left(
-\allowbreak 1\right) ^{k}\text{sech}^{l+k}2\theta }{l!k!}  \notag \\
&&\times \frac{\partial ^{k+l}}{\partial E^{k}\partial F\allowbreak ^{\ast l}%
}\frac{\partial ^{k+l}}{\partial E\allowbreak ^{\ast k}\allowbreak \partial
F^{l}\allowbreak }H_{n,n}\left( E,\frac{F^{\ast }}{\tanh 2\theta }\right)
H_{n,n}\left( E^{\ast },\frac{F}{\tanh 2\theta }\right) .  \label{38}
\end{eqnarray}%
Noticing the formula%
\begin{equation}
\frac{\partial ^{l+k}}{\partial \xi ^{l}\partial \eta ^{k}}H_{m,n}\left( \xi
,\eta \right) =\frac{m!n!}{\left( m-l\right) !\left( n-k\right) !}%
H_{m-l,n-k}\left( \xi ,\eta \right) ,  \label{39}
\end{equation}%
we have%
\begin{eqnarray}
W_{3}\left( \alpha \right) &=&\frac{n!^{2}e^{-2\left\vert \alpha \right\vert
^{2}\text{sech}2\theta }}{\pi \cosh 2\theta }\sum_{l,k=0}^{n}\frac{\left(
-\allowbreak 1\right) ^{k}\text{sech}^{l+k}2\theta \tanh ^{2\left(
n-l\right) }2\theta }{l!k!\left[ \left( n-l\right) !\left( n-k\right) !%
\right] ^{2}}  \notag \\
&&\times \left\vert H_{n-k,n-l}\left( E,\frac{F^{\ast }}{\tanh 2\theta }%
\right) \right\vert ^{2}.  \label{40}
\end{eqnarray}%
From Eq.(\ref{40}) one can see clearly that the WF of thermo number state is
a real number.

In particular, when $n=0$, noticing that $\tanh \theta =e^{-\frac{1}{2}%
\omega \beta },$ $\cosh ^{2}\theta =\frac{1}{1-e^{-\beta \omega }},\sinh
^{2}\theta =\frac{e^{-\beta \omega }}{1-e^{-\beta \omega }},$ Eq.(\ref{40})
reduces to the WF of thermo vacuum state $\left\vert 0(\beta )\right\rangle $
in Eq.(\ref{19})$.$ On the other hand, when $T\rightarrow 0,$(i.e., finite
temperature case reduces to zero temperature case) $e^{-\beta \omega
}\rightarrow e^{-\infty }\rightarrow 0,$ $\sinh \theta \rightarrow 0,$ $%
\cosh \theta \rightarrow 1,$ $E\rightarrow 2\alpha ,$ $\frac{F^{\ast }}{%
\tanh 2\theta }\rightarrow \alpha ^{\ast },$ and noticing Eq.(\ref{27}) and
the definition of two-variable Hermite polynomials [24,25],
\begin{equation}
H_{m,n}\left( \xi ,\kappa \right) =\sum_{l=0}^{\min (m,n)}\frac{m!n!\left(
-1\right) ^{l}\xi ^{m-l}\kappa ^{n-l}}{l!\left( n-l\right) !\left(
m-l\right) !},  \label{41}
\end{equation}%
which leads to $H_{n-k,0}\left( 2\alpha ,\alpha ^{\ast }\right) =\left(
2\alpha \right) ^{n-k},$ then Eq.(\ref{40}) becomes%
\begin{eqnarray}
W_{3}\left( \alpha \right)  &=&\frac{1}{\pi }e^{-2\left\vert \alpha
\right\vert ^{2}}\sum_{k=0}^{n}\frac{\left( -\allowbreak 1\right) ^{k}n!}{k!%
\left[ \left( n-k\right) !\right] ^{2}}\left\vert H_{n-k,0}\left( 2\alpha
,\alpha ^{\ast }\right) \right\vert ^{2}  \notag \\
&=&\frac{\left( -\allowbreak 1\right) ^{n}}{\pi }e^{-2\left\vert \alpha
\right\vert ^{2}}\sum_{k=0}^{n}\frac{n!}{k!\left[ \left( n-k\right) !\right]
^{2}}\left( -4\left\vert \alpha \right\vert ^{2}\right) ^{n-k}  \notag \\
&=&\frac{\left( -\allowbreak 1\right) ^{n}}{\pi }e^{-2\left\vert \alpha
\right\vert ^{2}}L_{n}(4\left\vert \alpha \right\vert ^{2}),  \label{42}
\end{eqnarray}%
which is just the WF of number state $\left\vert n\right\rangle $\ at zero
temperature.

In sum, by using TFD and Weyl ordered operators' order-invariance under
similar transformations, we present a new approach to deriving the exact
expressions of Wigner functions for thermo number state, photon subtracted
and added thermo vacuum state. These WF are related to the Gaussian-Laguerre
type functions, which are easily to be further analysed. The affection of
temperature to nonclassical behaviour of the fields is manifestly shown. For
discussions about the decoherence at finite temperature, we refer to [30,31].

\textbf{Appendix A} Checking Eq.(\ref{28})

In fact, in original Fock space, the photon-subtracted thermo state is
expressed as [20]%
\begin{equation}
\rho _{1}=C_{1}\text{Tr}_{\tilde{a}}\left[ a^{n}\left\vert 0(\beta
)\right\rangle \left\langle 0(\beta )\right\vert a^{\dag n}\right]
=C_{1}a^{n}\rho _{c}a^{\dag n},  \tag{A1}
\end{equation}%
where $\rho _{c}$ is the thermo state%
\begin{equation}
\rho _{c}=\sum_{l=0}^{\infty }\frac{n_{c}^{l}}{\left( n_{c}+1\right) ^{l+1}}%
\left\vert l\right\rangle \left\langle l\right\vert =\frac{1}{n_{c}+1}%
e^{a^{\dag }a\ln \frac{n_{c}}{n_{c}+1}},\text{ }n_{c}=\sinh ^{2}\theta .
\tag{A2}
\end{equation}%
Using the the coherent state representation of Wigner operator [26],
\begin{equation}
\Delta \left( \alpha \right) =e^{2\left\vert \alpha \right\vert ^{2}}\int
\frac{\mathtt{d}^{2}z}{\pi ^{2}}\left\vert z\right\rangle \left\langle
-z\right\vert \exp \left[ -2\left( z\alpha ^{\ast }-z^{\ast }\alpha \right) %
\right] ,  \tag{A3}
\end{equation}%
where $\left\vert z\right\rangle $ is the coherent state [27,28], we have
\begin{align}
W_{1}\left( \alpha \right) & =\text{Tr}\left( \Delta \left( \alpha \right)
\rho _{1}\right)   \notag \\
& =\frac{C_{1}e^{2\left\vert \alpha \right\vert ^{2}}}{n_{c}+1}\int \frac{%
\mathtt{d}^{2}z}{\pi ^{2}}\left\langle -z\right\vert a^{n}e^{a^{\dag }a\ln
\frac{n_{c}}{n_{c}+1}}a^{\dag n}\left\vert z\right\rangle \exp \left[
-2\left( z\alpha ^{\ast }-z^{\ast }\alpha \right) \right] .  \tag{A4}
\end{align}%
Note that
\begin{equation}
e^{a^{\dag }a\ln \frac{n_{c}}{n_{c}+1}}a^{\dag n}e^{-a^{\dag }a\ln \frac{%
n_{c}}{n_{c}+1}}=\frac{n_{c}^{n}}{\left( n_{c}+1\right) ^{n}}a^{\dag n},
\tag{A5}
\end{equation}%
and
\begin{equation}
e^{a^{\dag }a\ln \frac{n_{c}}{n_{c}+1}}\left\vert z\right\rangle =e^{-\frac{%
2n_{c}+1}{2\left( n_{c}+1\right) ^{2}}}\left\vert \frac{n_{c}z}{n_{c}+1}%
\right\rangle ,  \tag{A6}
\end{equation}%
Eq.(A4) can be rewritten as%
\begin{align}
W_{1}\left( \alpha \right) & =\frac{n_{c}^{n}C_{1}e^{2\left\vert \alpha
\right\vert ^{2}}}{\left( n_{c}+1\right) ^{n+1}}\int \frac{d^{2}z}{\pi ^{2}}%
\left\langle -z\right\vert a^{n}a^{\dag n}\left\vert \frac{n_{c}z}{n_{c}+1}%
\right\rangle   \notag \\
& \times \exp \left[ -\frac{2n_{c}+1}{2\left( n_{c}+1\right) ^{2}}\left\vert
z\right\vert ^{2}-2\left( z\alpha ^{\ast }-z^{\ast }\alpha \right) \right] .
\tag{A7}
\end{align}%
Further using the operator identity [29]%
\begin{equation}
a^{n}a^{\dag n}=\left( -1\right) ^{n}\colon H_{n,n}\left( ia^{\dag
},ia\right) \colon ,  \tag{A8}
\end{equation}%
where $H_{m,n}\left( x,y\right) $ is the two-variable Hermite polynomials,
whose generating function is
\begin{equation}
H_{m,n}\left( x,y\right) =\left. \frac{\partial ^{m+n}}{\partial
t^{m}\partial t^{\prime n}}\exp \left[ -tt^{\prime }+tx+t^{\prime }y\right]
\right\vert _{t=t^{\prime }=0},  \tag{A9}
\end{equation}%
we have%
\begin{align}
W_{1}\left( \alpha \right) & =\frac{\left( -1\right) ^{n}e^{2\left\vert
\alpha \right\vert ^{2}}}{n!\left( n_{c}+1\right) ^{n+1}}\frac{\partial ^{2n}%
}{\partial t^{n}\partial \tau ^{n}}e^{-t\tau }  \notag \\
& \times \int \frac{\mathtt{d}^{2}z}{\pi ^{2}}\exp \left\{ -\frac{2n_{c}+1}{%
n_{c}+1}\left\vert z\right\vert ^{2}\right. \left. +\left( \frac{i\tau n_{c}%
}{n_{c}+1}-2\alpha ^{\ast }\right) z+\left( 2\alpha -it\right) z^{\ast
}\right\} _{t=\tau =0}  \notag \\
& =\frac{\left( -1\right) ^{n}}{n!\pi \left( n_{c}+1\right) ^{n}}\frac{%
e^{2\left\vert \alpha \right\vert ^{2}}}{2n_{c}+1}\frac{\partial ^{2n}}{%
\partial t^{n}\partial \tau ^{n}}\exp \left[ -t\tau \right]   \notag \\
& \exp \left[ \frac{n_{c}+1}{2n_{c}+1}\left( \frac{i\tau n_{c}}{n_{c}+1}%
-2\alpha ^{\ast }\right) \left( 2\alpha -it\right) \right] _{t=\tau =0}
\notag \\
& =\frac{\left( -1\right) ^{n}}{n!\pi \left( n_{c}+1\right) ^{n}}\frac{e^{-%
\frac{2\left\vert \alpha \right\vert ^{2}}{2n_{c}+1}}}{2n_{c}+1}\frac{%
\partial ^{2n}}{\partial t^{n}\partial \tau ^{n}}\exp \left\{ -\frac{n_{c}+1%
}{2n_{c}+1}t\tau \right.   \notag \\
& +\left. 2i\alpha ^{\ast }\frac{n_{c}+1}{2n_{c}+1}t+2i\alpha \frac{n_{c}}{%
2n_{c}+1}\tau \right\} _{t=\tau =0}  \notag \\
& =\frac{e^{-\frac{2\left\vert \alpha \right\vert ^{2}}{2n_{c}+1}}}{\left(
2n_{c}+1\right) ^{n+1}}\frac{\left( -1\right) ^{n}}{n!\pi }H_{n,n}\left(
\frac{2in_{c}\alpha }{\sqrt{\left( 2n_{c}+1\right) \left( n_{c}+1\right) }}%
,2i\sqrt{\frac{n_{c}+1}{2n_{c}+1}}\alpha ^{\ast }\right) ,  \tag{A10}
\end{align}%
then using the relation
\begin{equation}
\frac{\left( -1\right) ^{n}}{n!}H_{n,n}\left( x,y\right) =L_{n}\left(
xy\right) ,  \tag{A11}
\end{equation}%
and noticing that $n_{c}=\sinh ^{2}\theta ,$ $2n_{c}+1=\cosh 2\theta ,$
Eq.(A10) can be put into
\begin{equation}
W_{1}\left( \alpha \right) =\frac{e^{-\frac{2\left\vert \alpha \right\vert
^{2}}{2n_{c}+1}}}{\pi \left( 2n_{c}+1\right) ^{n+1}}L_{n}\left( -\frac{%
4n_{c}\left\vert \alpha \right\vert ^{2}}{2n_{c}+1}\right) ,  \tag{A12}
\end{equation}%
which is just the Eq.(\ref{28}). Thus we have checked the result using a new
appraoch.

\bigskip

\bigskip

\bigskip


\begin{thebibliography}{99}
\bibitem{1} Parigi V, Zavatta A, Kim M S and Bellini M 2007 Science \textbf{%
317} 1890

\bibitem{2} Boyd R W, Chan K W and O'Sullivan M N 2007 Science \textbf{317}
1874

\bibitem{3} Wenger J, Tualle-Brouri R and Grangier P 2004 Phys. Rev. Lett.
\textbf{92} 153601

\bibitem{4} Zavatta A, Viciani S and Bellini M 2004 Science 306 660

\bibitem{5} Ourjoumtsev A, Dantan A, Tualle-Brouri R and Grangier P 2007
Phys. Rev. Lett. \textbf{98} 030502

\bibitem{6} Ourjoumtsev A, Dantan A, Tualle-Brouri R and Grangier P 2006
Phys. Rev. Lett. \textbf{96} 213601

\bibitem{7} Biswas A and Agarwal G S 2007 Phys. Rev. A \textbf{75} 032104

\bibitem{8} Li-yun Hu and Hong-yi Fan 2008 J. Opt. Soc. Am. B \textbf{25}
1955

\bibitem{9} Wigner E, 1932 Phys. Rev. \textbf{40} 749

\bibitem{10} Wolfgang P. Schleich, Quantum Optics in Phase Space, Wiley-VCH,
Birlin, 2001

\bibitem{11} Kim M S and Bu\v{z}ek V, 1992 Phys. Rev. A \textbf{46}
4239-4251.

\bibitem{12} Jeong H, Lund A P, and Ralph T C, 2005 Phys. Rev. A \textbf{72}
013801; Jeong H, Lee J and Nha H, 2008 J. Opt. Soc. Am. B \textbf{25} 1025

\bibitem{13} Hong-yi Fan, 1992 J. Phys. A \textbf{25} 3443; Hong-yi Fan,
2008 Ann. Phys. \textbf{323} 500

\bibitem{14} Hong-yi Fan, 1997 Mod. Phys. Lett. A \textbf{12} 2325; 2000
Mod. Phys. Lett. A \textbf{15} 2297

\bibitem{15} Hong-yi Fan and Yue Fan, 1998 Mod. Phys. Lett. A \textbf{13}
433; 2002 Int. J. Mod. Phys. A \textbf{17} 701

\bibitem{16} Y. Takahashi and Umezawa H, 1975 Collecive Phenomena \textbf{2}
55

\bibitem{17} Memorial Issue for Umezawa H, 1996 \textit{Int. J. Mod. Phys. B}
\textbf{10} 1695 memorial issue and references therein.

\bibitem{18} H. Umezawa, Advanced Field Theory -- Micro, Macro, and Thermal
Physics (AIP 1993).

\bibitem{19} R. R. Puri, \textit{Mathematical Methods of Quantum Optics}
(Springer-Verlag, Berlin, 2001), Appendix A.

\bibitem{20} Agarwal G S 1992 Phys. Rev. A \textbf{45} 1787

\bibitem{21} Hong-yi Fan, et. al., 2006 Ann. Phys. \textbf{321} 480

\bibitem{22} Magnus W et. al., Formulas and Theorems for the Special
Functions of Mathematical Physics, 3rd Ed., Springer Verlag. 1966

\bibitem{23} Agarwal G S and Tara K 1992 Phys. Rev. A \textbf{46} 485

\bibitem{24} W\"{u}nsche A, 2001 \textit{J. Computational and Appl. Math.}
\textbf{133} 665

\bibitem{25} W\"{u}nsche A, 2000 \textit{J . Phys. A: Math. and Gen. }%
\textbf{33} 1603

\bibitem{26} Hong-yi Fan, 1987 Phys. Lett. A \textbf{124} 303

\bibitem{27} Glauber R J, 1963 Phys. Rev. \textbf{130} 2529-2539; 1963 Phys.
Rev. \textbf{131} 2766-2788

\bibitem{28} Klauder J R and Skargerstam B S, Coherent States (World
Scientific, Singapore, 1985).

\bibitem{29} Hong-yi Fan, 2004 Commun. Theor. Phys. \textbf{42} 339

\bibitem{30} Hong-yi Fan and Li-yun Hu, 2008 Mod. Phys. Lett. B \textbf{22}
2435; Hong-yi Fan and Hai-liang Lu, 2007 Mod. Phys. Lett. B \textbf{21} 183

\bibitem{31} Hong-yi Fan and Li-yun Hu, 2008 Opt. Commun. \textbf{281} 5571
\end{thebibliography}
\end{document}